\input{aipcheck}

\documentclass[
    ,final            
  ]
  {aipproc}

\layoutstyle{6x9}

\begin{document}

\title{The effective field theory treatment of quantum gravity}


\author{John F. Donoghue}{
  address={Department of Physics, University of Massachusetts, Amherst, MA 01003, U.S.A.}
}

\begin{abstract}
This is a pedagogical introduction to the treatment of quantum
general relativity as an effective field theory. It starts with an
overview of the methods of effective field theory and includes an
explicit example. Quantum general relativity matches this framework
and I discuss gravitational examples as well as the limits of the effective
field theory. I also discuss the insights from effective field theory on the gravitational
effects on running couplings in the perturbative regime.
\end{abstract}

\maketitle


\section{Introduction}

Several years ago, I wrote some lectures\cite{Donoghue:1995cz} on the application of effective field theory (EFT)
to general relativity. Because those lectures were given at an advanced school on effective field theory, most
of the students were well versed about the effective field theory side, and the  point was to show how general
relativity matches standard effective field theory practice. In contrast, the present lectures were delivered at
a school where the students primarily knew general relativity, but knew less about effective field theory. So
the written version here will focus more on the effective field theory side, and can be viewed as complementary
to previous lectures. There will be some repetition and updating of portions of the previous manuscript, but the
two can potentially be used together\footnote{The reader can also consult my original paper on the subject\cite{Donoghue:1994dn}
or the review by Cliff Burgess\cite{Burgess:2003jk}, and of course Steven
Weinberg's thoughts on effective field theory\cite{Weinberg:2009bg} are always interesting.}.

There has been a great deal of good work combining general relativity and field theory. In the
early days the focus was on the high energy behavior of the theory, especially on the divergences.
However, this is the portion of gravity theory that we are most unsure about - the conventional
expectation is that gravity needs to be modified beyond the Planck scale. What effective field
theory does is to shift the focus to low energy where we reliably know the content of general
relativity. It provides a well-defined framework for organizing quantum calculations and
understanding which aspects are reliable and which are not. In this sense, it may provide the
maximum content of quantum general relativity until/unless we are able to solve the mystery of Planck scale physics.

Effective field theory has added something important to the understanding of quantum gravity.
One can find thousands of statements in the literature to the effect that ``general relativity and
quantum mechanics are incompatible''. These are completely outdated and no longer relevant. Effective
field theory shows that general relativity and quantum mechanics work together perfectly normally
over a range of scales and curvatures, including those relevant for the world that we see around us.
However, effective field theories are only valid over some range of scales. General relativity
certainly does have problematic issues at extreme scales. There are important problems which
the effective field theory does not solve because they are beyond its range of validity. However,
this means that the issue of quantum gravity is not what we thought it to be. Rather than a fundamental
incompatibility of quantum mechanics and gravity, we are in the more familiar situation of needing a
more complete theory beyond the range of their combined applicability. The usual marriage
of general relativity and quantum mechanics is fine at ordinary energies, but we now seek to uncover
the modifications that must be present in more extreme conditions. This is the modern view of the
problem of quantum gravity, and it represents progress over the outdated view of the past.

\section{Effective field theory in general}

Let us start by asking how {\em any} quantum mechanical calculation can be reliable.
Quantum perturbation theory instructs us to sum over {\em all} intermediate states of {\em all} energies.
However, because physics is an experimental science, we do not know all the states that exist at high
energy and we do not know what the interactions are there either. So why doesn't this lack of knowledge
get in the way of us making good predictions? It is not because energy denominators cut off the high
energy portion - in fact phase space favors high energy. It is not because the
transition matrix elements are small. In fact, one can argue that all processes are sensitive to
the highest energy at some order in perturbation theory.

The answer is related to the uncertainty principle. The effects of the highest energies correspond
to such short distances that they are effectively local when viewed at low energy. As such they are
identical to some local term in the Lagrangian that we use to define our low energy theory. These terms
in the Lagrangian come with coupling constants or masses as coefficients, and then the effect of the highest
energy goes into measured value of these parameters.
This is a well-known story for divergences, but is true for finite effects also,
and would be important for quantum predictions even if the ultimate theory had no divergences.

In contrast, low energy effects of massless (or very light) particles are not local. Examples include
the photon propagator or two photon exchange potentials. Intermediate states that go on-shell clearly
propagate long distances. Sometimes loop diagrams of light particles can have both short and long distance
contributions within the same diagram. We will see that we can separate these, because the short distant
part looks like a local effect and we can catalog all the local effects. The long distance portions of
processes can be separated from the short distance physics.

Effective field theory is then the procedure for describing the long-distance physics of the light
particles that are active at low energy. It can be applied usefully in both cases where we know the
full theory or when we do not. In both situations the effects of heavy particles and interactions
(known or unknown) are described by a local effective Lagrangian, and we treat the light particles
with a full field theoretic treatment, including loops, renormalization, etc.

\section{Constructing an explicit effective field theory - the sigma model}

Perhaps the best way to understand the procedures of effective field theory is to make an explicit construction
of one from a renormalizable theory\footnote{The treatment of this section follows that of
our book\cite{Donoghue:1992dd}, to which the reader is referred for more information.}. This can be done
using the linear sigma model. Moreover the resulting effective field theory has a form with many similarities to general relativity.

Let us start with the bosonic sector of the linear sigma model\footnote{The full linear sigma
model also includes coupling to a fermion doublet, but that aspect is not needed for our purpose.}
\begin{eqnarray}
{\cal L} &=&  \frac12
\partial_\mu \vec{\pi}\cdot \partial^\mu \vec{\pi} + \frac12
\partial_\mu\sigma\partial^\mu \sigma \nonumber \\
&+& {1\over 2} \mu^2 \left( \sigma^2 + {\vec{\pi}}^2\right) -
{\lambda\over 4} \left( \sigma^2 + \vec{\pi}^2\right)^2 \ \ .
\end{eqnarray}
with four fields $\sigma,~ \pi_1,~\pi_2,~\pi_3 $ and an obvious invariance of rotations
among these fields. This will be referred to as "the full theory". The invariance can be
described by $SU(2)_l\times SU(2)_R$ symmetry, with the transformation being
\begin{equation}
\sigma + i \vec{\tau} \cdot \vec{\pi} \to V_L \left(\sigma + i \vec{\tau }\cdot \vec{\pi} \right)V^\dagger_R
\end{equation}
where $V_{L,R}$ are $2\times 2$ elements of $SU(2)_{L,R}$, and $\tau^i$ are the Pauli $SU(2)$ matrices.

This theory is the well-known example of spontaneous symmetry breaking with the ground state
being obtained for $<\sigma >=v =\sqrt{\mu^2/\lambda}$. The $\sigma$ field picks up a
mass $m_\sigma^2 =2\mu^2$, and the $\vec{ \pi}$ fields are massless Goldstone bosons.
After the shift $\sigma =v + \tilde{\sigma}$, the theory becomes
\begin{eqnarray}
{\cal L} &=&  \frac12
\partial_\mu \vec{\pi}\cdot \partial^\mu \vec{\pi} + \frac12\left[
\partial_\mu\tilde{\sigma}\partial^\mu \tilde{\sigma} +m_\sigma^2 \tilde{\sigma}^2\right]\nonumber \\
&-& \lambda v\tilde{ \sigma}(\tilde{\sigma}^2 + \vec{\pi}^2 )-
{\lambda\over 4} \left( \tilde{\sigma}^2 + \vec{\pi}^2\right)^2 \ \
\label{shifted}
\end{eqnarray}

The only dimensional parameter in this theory is $v$ or equivalently $\mu= \sqrt{\lambda} v$ or $m_\sigma =\sqrt{2}\mu$,
all of which carry the dimension of a mass. For readers concerned with gravity, you should think of
this as ``the Planck mass'' of the theory. Consider it to be very large, well beyond any energy that
one can reach with using just the massless $\vec{\pi}$ fields.

At low energy, well below ``the Planck mass'', all that you will see are the pions and you
will seek a Lagrangian describing their interactions without involving the $\sigma$. This is not
obvious looking at Eq. \ref{shifted}, but can be done. That result is
\begin{equation}
{\cal L}_{\rm eff} = {v^2\over 4}
{\rm Tr} \left(\partial_\mu U\partial^\mu U^\dagger\right)
\label{Leff}
\end{equation}
where
\begin{equation}
U = \exp [ \frac{i\vec{\tau}\cdot \vec{\pi}}{v}] \ \ .
\label{U}
\end{equation}
This of course looks very different. It is a ``non-renormalizable'' Lagrangian where the
exponential generates non-linear interactions to all orders of the pion field. It also
comes with two derivatives acting on the fields, so that all matrix elements are proportional
to the energy or momentum of the fields squared. In contrast, the original sigma model has
non-derivative interactions such as the $\lambda {\mathbf \pi}^4$ coupling. As we will see
in more detail, these features of the effective theory are shared with general relativity.

\subsection{Example of the equivalence}

Before I describe the derivation of the effective theory, let us look at an explicit calculation
in order to see the nature of the approximation involved. If we calculate a process
such as $\pi^+\pi^0\to \pi^+\pi^0$ scattering, in the full theory one finds both a
direct pion coupling and also a diagram that involves $\sigma $ exchange which arises
from the trilinear coupling in Eq. \ref{shifted}, as shown in Fig. 1.

\begin{figure}
$\begin{array} {c@{\hspace{0.01 in}}c} \multicolumn{1}{l}{} &
\multicolumn{1}{l}{} \\
{\resizebox{2.5in}{!}{\includegraphics{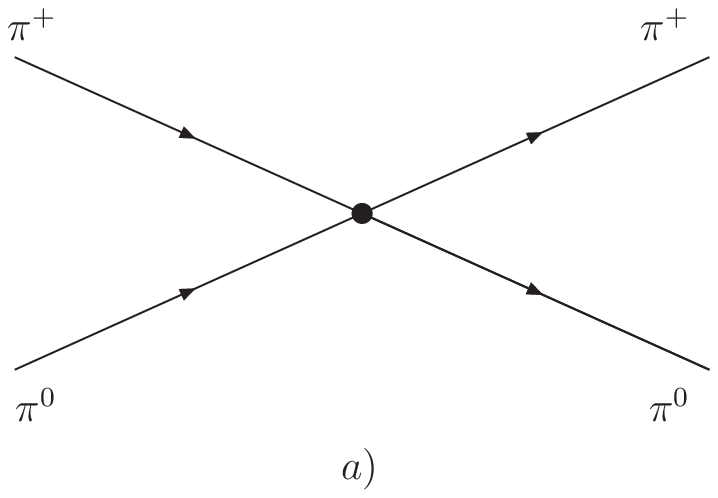}}}&
~~~~~~~~{\resizebox{2.5in}{!}{\includegraphics{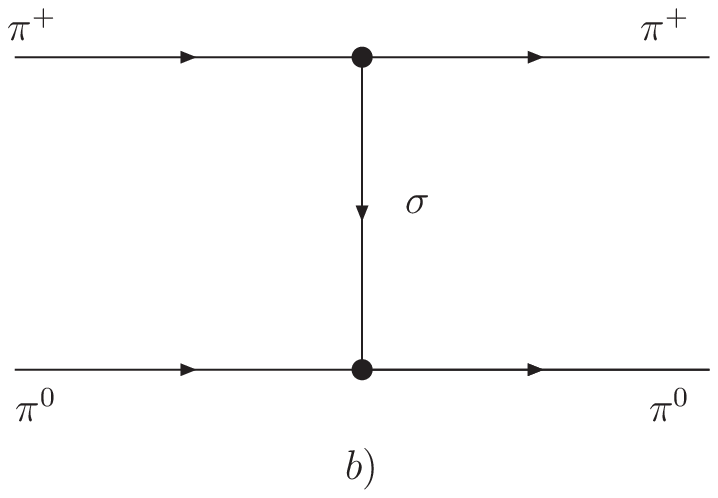}}} \\
\end{array}$
\caption{Diagrams for $\pi -\pi$ scattering}

\end{figure}

This results in a cancelation such that the matrix element is
\begin{eqnarray}
i{\cal M}_{\pi^+ \pi^0\to \pi^+\pi^0} = - 2i\lambda
+ \left( -
2i\lambda v\right)^2 {i\over q^2 - m^2_\sigma}
= - 2i\lambda\left[ 1 + {2\lambda v^2\over q^2 - 2\lambda v^2}\right] \\
=  +i\left[\frac{q^2}{v^2} + \frac{q^4}{v^2m_\sigma^2}+ \ldots\right] \ \ .
\label{full tree}
\end{eqnarray}
Here $q^2 =(p_+-p_+')^2 =t$ is the momentum transfer\footnote{For future use, the reader is reminded of the
Mandelstam variables $s= (p_++p_0)^2,~t=(p_+-p_+')^2=q^2,~ u=(p_0-p_+')^2$.} and I remind the reader that $m_\sigma^2 =2\lambda v^2$.
Despite the full theory
having only polynomial interactions without derivatives, the result ends up proportional to $q^2$. Here
we also see the nature of the energy expansion. The correction to the leading term is
suppressed by two powers of the sigma mass (i.e. ``the Planck mass'').

In the effective theory one proceeds by expanding the exponential in Eq. \ref{Leff} to order $\pi^4$ and
taking the matrix element. Because the Lagrangian has two derivatives, the result will automatically be of order $q^2$. We find
\begin{equation}
i{\cal M}_{\pi^+ \pi^0\to \pi^+\pi^0} = {iq^2\over v^2}
\end{equation}
which of course is exactly the same as the first term in the scattering amplitude of the
full theory. Moreover, exploration of other matrix elements will show that {\em all} amplitudes
will agree for the full theory and the effective theory at this order in the energy expansion.
This is a very non-trivial fact. We will see how one obtains the next order term (and more) soon.

\subsection{Construction of the effective theory}

Lets build the effective theory starting with the original linear sigma model.
Instead of the redefinition of the field $\sigma = v +\tilde{\sigma}$, let us consider the renaming
\begin{equation}
\sigma + i \vec{\tau} \cdot \vec{\pi} = (v+ \sigma') \exp [ \frac{i\vec{\tau}\cdot\vec{\pi}'}{v}] = v +\sigma' +i\vec{\tau} \cdot \vec{\pi}' +\ldots
\end{equation}
This change of fields from $(\sigma, \vec{\pi})$ to $(\sigma', \vec{\pi'})$ is simply a renaming of the fields
which by general principles of field theory will not change the physical amplitudes. Using the new
fields (although quickly dropping the primes for notational convenience) the full original linear sigma model can be rewritten without
any approximation as
\begin{equation}
{\cal L}_{\rm eff} = \frac14 (v+\sigma)^2 {\rm Tr} \left(\partial_\mu U\partial^\mu U^\dagger\right) + {\cal L}(\sigma)
\label{alternate}
\end{equation}
where $U$ is the exponential of the pion fields of Eq. \ref{U} and
\begin{equation}
{\cal L}(\sigma) =  \frac12\left[
\partial_\mu{\sigma}\partial^\mu {\sigma} +m_\sigma^2 {\sigma}^2\right]
- \lambda v \sigma^3  -\frac{\lambda}{4} \sigma^4   \ \ .
\end{equation}
This alternate form also describes a heavy field $\sigma$ with a set of self interactions and
with couplings to the massless pions. The symmetry of the original
theory $\sigma + i \vec{\tau} \cdot \vec{\pi} \to V_L \left(\sigma + i \vec{\tau} \cdot \vec{\pi} \right)V^\dagger_R$ is still
manifest as $U \to  V_L U V^\dagger_R,~~\sigma ' \to \sigma '$.
This form {\em looks} non-renormalizable, but is not - it is really just the same theory
as the original form.

To transition to the effective field theory, we note that the $\sigma$ lives at "the Planck scale"
and therefore is inaccessible at low energy. Only virtual effects of the $\sigma$ can have any influence
on low energy physics. By inspection we see that the coupling of the heavy particle to the pions involves
two derivatives (i.e. powers of the energy), so that exchange of the $\sigma$ will involve at least four
powers of the energy. (We will treat loops soon.) So it appears fairly obvious that the leading effect - accurate to two derivative order -
is simply to neglect the exchange of the $\sigma$, directly leading to the effective Lagrangian of Eq. \ref{Leff}.

To confirm and extend this, it is useful to calculate explicitly the effect of $\sigma$ exchange.
The effect of this is pictured schematically in Fig 2, where the $\times$ represents
the "current" $\sim v {\rm Tr} \left(\partial_\mu U\partial^\mu U^\dagger\right) /2$. This diagram
yields the modification to the effective Lagrangian
\begin{equation}
{\cal L}_{\rm eff} = {v^2\over 4}
{\rm Tr} \left(\partial_\mu U\partial^\mu U^\dagger\right)+ \frac{v^2}{8 m_\sigma^2}[{\rm Tr} \left(\partial_\mu U\partial^\mu U^\dagger\right)]^2
\label{L4}
\end{equation}
As expected, the modification involves four derivatives. Moreover, if we take the
matrix element for the scattering amplitude calculated in the previous section,
we recover exactly the $q^4$ term found in the full theory, Eq. \ref{full tree}. Using the effective Lagrangian framework
we are able to match, order by order in the energy expansion, the results of the full theory.
\begin{figure}
  \includegraphics[height=.2\textheight]{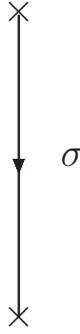}
  \caption{The tree-level effect of the exchange of a heavy scalar, $\sigma$. The $\times$ represents a vertex involving pions, as described in the text.}
\end{figure}

\subsection{Loops and renormalization}

The real power of effective field theory comes when we treat it as a full field theory - including loop diagrams -
rather than just a set of effective Lagrangians. The idea is that we "integrate out" the heavy field, leaving a
result depending on the light fields only. In path integral notation this means
\begin{eqnarray}
{\cal Z } &=& \int  [d \pi][d\sigma ] e^{i \int d^4x {\cal L}(\sigma , \pi) }    \nonumber \\
    &=& \int [d{\pi}] e^{i \int d^4x {\cal L}_{eff}( \pi) }
\end{eqnarray}
In practice, it means that we treat loop diagrams involving the $\sigma$ as well as tree diagrams.
But in loops also, the heavy particle does not propagate far, so
even the effects of heavy particle in loops can be represented by local Lagrangians.

The key to determining the low energy effective field theory is to {\em match} the predictions of
the full theory to that of the effective theory. To be sure, the effective field theory has an
incorrect high energy behavior. This can be seen even within specific diagrams. For example consider the diagram of
Fig. 3a which occurs within the full theory. This diagram is finite, although pretty complicated. When treated as an
effective theory we approximate the sigma propagator as a constant, shrinking the effect of sigma exchange to a local
vertex. This leads to the diagram of Fig. 3b, which is clearly divergent\footnote{The Taylor expansion of the propagator
will lead to increasingly divergent terms also.}. So new divergences will occur in the effective field theory
treatment which are not present in the full theory.

\begin{figure}
$\begin{array} {c@{\hspace{0.01 in}}c} \multicolumn{1}{l}{} &
\multicolumn{1}{l}{} \\
{\resizebox{2.5in}{!}{\includegraphics{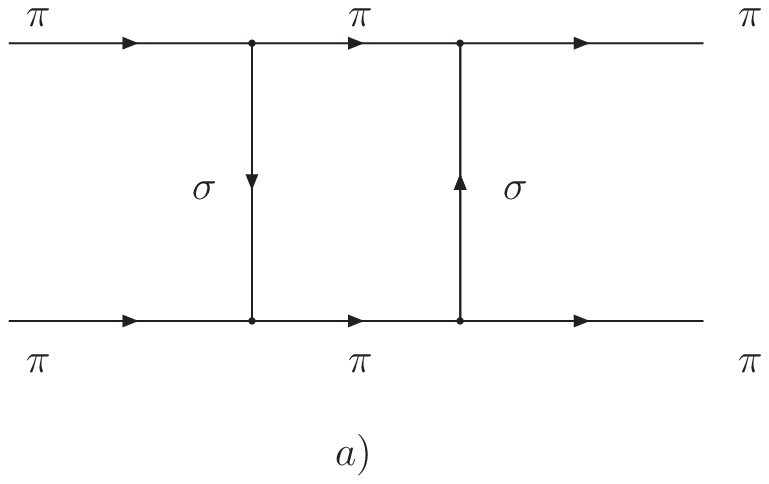}}}&
~~~~~~~~{\resizebox{2.5in}{!}{\includegraphics{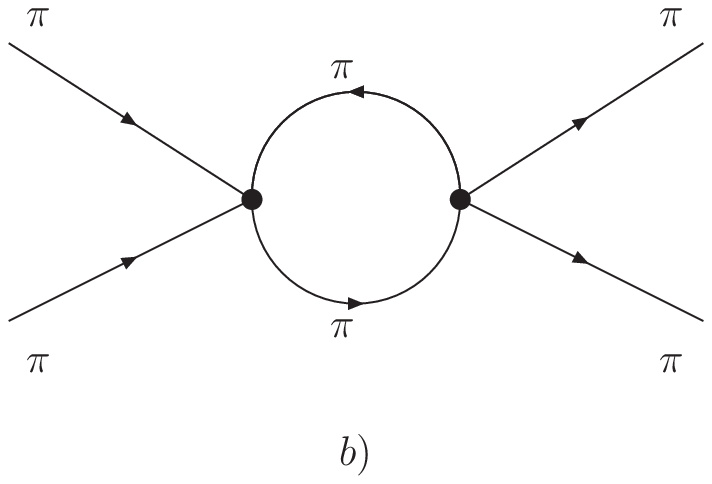}}} \\
\end{array}$
\caption{a) A finite box diagram which occurs in the full theory, b) A bubble diagram which occurs
  in the effective theory in the situation when the
  propagator of the heavy $\sigma$ has been shrunken to a point. }

\end{figure}

However, the low energy behavior of Fig. 3a and Fig 3b will be similar. When the loop momentum is
very small, the approximation of sigma exchange as a local vertex will be appropriate, and the pions will propagate long
distances. This will be manifest in non-analytic behaviors such as $\ln (-s)$ dependence at low energy\footnote{The $-i\pi$ which follows
from $\ln (-s)= \ln s -i\pi$ for $s>0$ accounts for the discontinuity from on-shell intermediate states. }. If the reader
has not yet struggled with the complexity of box diagrams, it would be a good exercise to look up the
box diagram for Fig 3a\cite{Denner:1991qq} and verify that the $\ln (-s)$ behavior is identical
(in the low energy limit) to that of the much simpler diagram Fig 3b.

In order to deal with the general divergences of the effective theory we need to
have the most general local Lagrangian. With the symmetry of the theory being $U\to V_LUV^\dagger_R$ the
general form with up to four derivatives is
\begin{equation}
{\cal L}_{eff} = \frac{M_P^2}{4} ~{\rm Tr}[\partial_\mu U \partial^\mu U^\dagger] + {\cal \ell}_1\left({\rm Tr}[\partial_\mu U \partial^\mu
   U^\dagger] \right)^2 +{\cal \ell}_2{\rm Tr}[\partial_\mu U \partial_\nu U^\dagger]  {\rm Tr}[\partial^\mu U \partial^\nu U^\dagger]
\end{equation}
where ${\cal \ell}_i$ are constant coefficients. We saw at tree level that
\begin{equation}
{\cal \ell}_1 =\frac{v^2}{8m_\sigma^2}~,~~~~ {\cal \ell}_2 =0
\end{equation}
The divergences of the effective theory will be local and will go into the renormalization of these coefficients.

The comparison of the full theory and the effective theory can be carried out directly
for the reaction $\pi^+ \pi^0\to \pi^+\pi^0$ . The dimensionally
regularized result for the full theory is quite complicated - it is given in \cite{Manohar:2008tc}.
However, the expansion of the full theory at low energy in terms of renormalized parameters is relatively simple\cite{GL}
\begin{eqnarray}
{\cal M}_{full}&=& {t\over v^2} +\left[{1\over m_\sigma^2 v^2}-{11\over 96\pi^2v^4}\right]t^2 \cr
&-&{1\over 144\pi^2 v^4}[s(s-u) +u(u-s)] \cr
&-&{1\over 96\pi^2 v^4}\left[3t^2 \ln {-t\over m_\sigma^2}+s(s-u)\ln {-s\over m_\sigma^2}+u(u-s)\ln {-u\over m_\sigma^2}\right]
\end{eqnarray}
One calculates the same reaction in the effective theory, which clearly does not know about
the existence of the $\sigma$. The result \cite{GL, Lehmann:1972kv} has a very similar form,
\begin{eqnarray}
{\cal M}_{eff}&=& {t\over v^2} +\left[8\ell_1^r+2\ell_2^r+{5\over 192\pi^2}\right]{t^2\over v^4} \\
&+&\left[2\ell_2^r+{7\over 576\pi^2}\right][s(s-u) +u(u-s)]/v^4 \\
&-&{1\over 96\pi^2 v^4}\left[3t^2 \ln {-t\over \mu^2}+s(s-u)\ln {-s\over \mu^2}+u(u-s)\ln {-u\over \mu^2}\right]
\end{eqnarray}
where we have defined the renormalized parameters
\begin{eqnarray}
\ell_1^r &=&\ell_1  +{1\over 384\pi^2 }\left[{2\over 4-d} - \gamma + \ln 4\pi \right] \\
\ell_2^r &=& \ell_2 +{1\over 192\pi^2 }\left[{2\over 4-d} - \gamma + \ln 4\pi \right]
\end{eqnarray}
At this stage we can match the two theories, providing identical scattering amplitudes to this order, through the choice
\begin{eqnarray}
\ell_1^r &=&  {v^2\over 8 m_\sigma^2} +{1\over 384\pi^2}\left[\ln {m_\sigma^2\over \mu^2} -{35\over 6}\right] \\
\ell_2^r &=& {1\over 192\pi^2 }\left[\ln {m_\sigma^2\over \mu^2} -{11\over 6}\right]
\end{eqnarray}
Loops have modified the result of tree level matching by a finite amount. We have not only obtained a more precise matching,
we also have generated important kinematic dependence, particularly the logarithms, in the scattering amplitude. The logarithms
do not involve the chiral coefficients $\ell_i$ because the logs follow from the long-distance portion of loops while the chiral
coefficients are explicitly short-distance. This is part of the evidence for a separation of long and short distances.

We have seen that we can renormalize a "nonrenormalizable" theory. The divergences are local and can be absorbed into
parameters of a local Lagrangian.  Moreover, the predictions of the full theory can be reproduced even when
using only the light degrees of freedom, as long as one chooses the coefficients of the effective lagrangian
appropriately. This holds for {\it all} observables. This can be demonstrated using a
background field method \cite{GL, Donoghue:1992dd, GL2}. Once the matching is done, other processes can be
calculated using the effective theory without the need to match again for each process. The total effect of
the heavy particle, both tree diagrams
and loops, has been reduced to a few numbers in the Lagrangian, which we have deduced
from matching conditions to a given order in an expansion in the energy.

\subsection{Power counting and the energy expansion}

We saw that the result of the one loop calculation involved effects that carried two extra factors
of the momenta $q^2$. It is relatively easy to realize that this is a general result. The only dimensional
parameter in the effective theory is $v$, and one can count the powers of $v$ that enter into loop diagrams.
These enter in the denominator because the exponential is expanded in $\tau \cdot \pi/v$. In dimensional regularlization
there is no scale to provide any competing powers in the renormalization
procedure\footnote{There is an arbitrary scale  $\mu$ which enters only logarithmically.}. Therefore the
external momenta $q^2, ~s, ~t, ~u$ must enter to compensate for the powers of $1/v^n$, leading to an expansion in $q^2/v^2$ or
more appropriately $q^2/(4\pi v)^2$.

This energy expansion is a second key feature of EFT techniques. Predictions are ordered in an
expansion of powers of the light energy scales over the high energy scale, ``the Planck scale'',
of the full theory. For the chiral theory, Weinberg provided a compact theorem showing how higher
loop diagrams always lead to higher energy dependence\cite{Weinberg:1978kz}. A similar power counting occurs in general relativity.

The power counting allows an efficient approach to loops in gravity also because general
relativity has a power counting behavior similar to the chiral case described here.
The coupling constant of gravity, Newton's constant $G$, carries the dimension of the
inverse Planck mass squared, $G\sim 1/M_P^2$. This means that loops carry extra powers of $G$
(just like the extra factors of $1/v^2$ of the chiral case) and this is compensated for by
factors of the low energy scale in the numerator, leading to an efficient energy expansion.

\subsection{Effective field theory in action}

Lest the above description sound somewhat formal, I should point out that the chiral effective
field theory is actively and widely used in phenomenological applications for QCD. Of course,
the linear sigma model is not the same as QCD. However, if the mass of the up and down quarks
were zero, QCD would also have an exact $SU(2)_L  \times SU(2)_R$ symmetry which is dynamically
broken, with massless pions as the Goldstone bosons. The effective Lagrangian of the theory would
have the same symmetry as the sigma model and so the general structure of the effective Lagrangian
would be the same. The identification of the vector and axial currents allows the identification of $v$ with
the pion decay constant $v=F_\pi = 92.4$~MeV measured in the decay $\pi \to \mu \nu$. The chiral
coefficients $\ell_1,~\ell_2$ are relatively different from those of the sigma model. In QCD they are
difficult to calculate from first principles, but as measured by experiment they are of order $10^{-3}$ and
their relative sizes reveal the influence of the vector meson $\rho(770)$ rather than of a scalar $\sigma$\cite{Donoghue:1988ed}.

The chiral symmetry of QCD also has small explicit breaking from the up and down quark masses, leading to a non-zero pion mass.
This symmetry breaking can also be treated perturbatively as an element in the energy expansion. The result is chiral perturbation
theory, a lively interplay of energies, masses, loop diagrams, experimental measurements that makes for a fine demonstration of the
practical power of effective field theory. The reader is referred to our book \cite{Donoghue:1992dd} for a more complete
pedagogic development of the subject.

There also are very many more uses of effective field theory throughout physics. Some of these have slightly
different issues and techniques, and the above is not a complete introduction to all aspects of effective field
theory. However, all EFTs share the common features of 1) using only the active degrees of freedom and interactions
relevant for the energy that one is working at, and 2) organizing the results in some form of an energy expansion.
The example of the linear sigma model is particularly well suited for the introduction to the gravitational theory.

\subsection{Summary of general procedures}

In preparation for the discussion of the effective field theory for general relativity, let me summarize the
techniques for an EFT in the situation when we do not know the full theory. In contrast with the linear sigma
model, we then cannot match the EFT to the full theory. However EFT techniques can still be useful and predictive.

1) One starts by identifying the low energy particles and the symmetries governing their interaction.

2) This information allows one to construct the most general effective Lagrangian using only these particles
and consistent with the symmetry. This Lagrangian can be ordered in an energy expansion in terms of increasing dimensions of the
operators involved. In contrast with the stringent constraints of "renormalizable field theory", one allows operators with
dimension greater than four. Renormalizable theories can be thought of as a subclass of EFTs where the operators of
dimension greater than four are not (yet) needed.

3) The operator(s) of lowest dimension provide the leading interactions at low energy. One can quantize the
fields and identify propagators in the usual way.

4) These interactions can now be used in a full field theoretic treatment. Loop diagrams can be computed.
Because the divergences will be local they can be absorbed into the renormalization of terms in the local effective Lagrangian.

5) Because, by assumption, the full theory is not known, the coefficients of various terms in the effective
Lagrangian cannot be predicted. However, in principle they can be measured experimentally. In this case,
these coefficients are not predictions of the effective field theory treatment - they represent information
about the full theory. However, once measured they can be used in multiple processes.

6) Predictions can be made. The goal is to obtain predictions from long distance (low energy) physics. Because
we know that short distance physics is local, it is equivalent to the local terms in the Lagrangian, and
everything left over is predictive. This most visible involves nonanalytic terms in momentum space, such
as the logarithmic terms in the scattering amplitude above. These are obviously not from a local Lagrangian
and correspond to long distance propagation. The logs can also pick up imaginary parts when their arguments
are negative - the imaginary parts are indicative of on-shell intermediate states. Analytic terms,
such as the $s^2/16\pi^2 v^2$ terms in the scattering amplitude above, can also be predictive when
they can be separated from the Lagrangian parameters, such as $\ell_i$, through measurement of the parameters in other reactions.

\section{Gravity as an effective field theory}

In the case of gravity, we assume that there is {\em some} well defined full theory of gravity that yields
general relativity as the low energy limit. We do not need to know what it is, but we take as experimental
fact that it is well behaved, with no significant instabilities or run-away solutions when dealing with
low curvature situations such as our own. Other particles and interactions - the Standard Model - can be
added, but are not crucial to the EFT treatment. Basically, we assume that there is a limit of the full
theory that looks like the world that we see around us.

If the low energy limit is general relativity, we know that the relevant degrees of freedom are massless
gravitons, which are excitations of the metric. The symmetry is general coordinate invariance. In constructing
the energy expansion of the effective Lagrangian, one must pay attention to the number of derivatives. The connection
\begin{equation}
\Gamma^\lambda_{\alpha\beta} = \frac{g^{\lambda\sigma}}{2}\left[\partial_\alpha g_{\beta\sigma}+\partial_\beta g_{\alpha\sigma}
- \partial_\sigma g_{\alpha\beta}\right]
\end{equation}
has one derivative of the metric, while the curvatures such as
\begin{equation}
R_{\mu\nu} = \partial_\mu \Gamma^\lambda_{\nu\lambda}-\partial_\lambda\Gamma^\lambda_{\mu\nu} -\Gamma^\sigma_{\mu\lambda}\Gamma^\lambda_{\nu\sigma}
-\Gamma^\sigma_{\mu\nu}\Gamma^\lambda_{\lambda\sigma}
\end{equation}
have two. The various contractions of the Riemann tensor are coordinate invariant. These features
determine the nature of the energy expansion of the the action for general relativity
\begin{equation}\label{gravL}
S_{grav}=\int d^4x\, \sqrt{g}\, \left[\Lambda +{2\over{\kappa^2}}\, R+c_1\, R^2 +
c_2\, R_{\mu \nu} R^{\mu \nu} + \ldots + {\cal L}_{matter}\,\right]
\end{equation}
Here the terms have zero, two and four derivatives respectively.

Following our EFT script, we turn to experiment to determine the parameters of this
Lagrangian. The first term, the cosmological constant, appears to be non-zero but it
is so tiny that it is not relevant on ordinary scales. The EFT treatment does not say
anything novel about the smallness of the cosmological constant - it is treated simply
as an experimental fact. The next term is the Einstein action, with coefficient determined from Newton's constant $\kappa^2 =32\pi G$.
This is the usual starting place for a treatment of general relativity. The curvature squared terms
yield effects that are tiny on normal scales if the coefficients $c_{1,2}$ are of order unity. In
fact these are bounded by experiment \cite{Stelle:1977ry} to be less than $10^{+74}$ - the ridiculous
weakness of this constraint illustrates just how irrelevant these terms are for normal physics. So we
see that general coordinate invariance allows a simple energy expansion.

At times people worry that the presence of curvature squared terms in the action will lead to instabilities or pathological behavior.
Such potential problems have been shown to only occur at scales beyond the Planck scale \cite{Simon:1990ic} where yet
higher order terms are also equally important. This is not a flaw of the effective field theory, which holds only
below the Planck scale. Given the assumption of a well-behaved full theory of gravity, there is no aspect of the
effective theory that needs to display a pathology.

\subsection{Quantization and renormalization}

The quantization of general relativity is rather like that of Yang-Mills theory. There are subtle features connected
with the gauge invariance, so that only physical degrees of freedom count in loops. Feynman\cite{Feynman:1963ax}, and
then DeWitt\cite{DeWitt:1967ub}, did this successfully in the 1960's, introducing gauge fixing and then ghost fields to
cancel off the unphysical graviton states. The background field method employed by 'tHooft and Veltman\cite{'tHooft:1974bx}
was also a beautiful step forward. It allows the expansion about a background metric ($\bar{g}_{\mu\nu}$) and explicitly
preserves the symmetries of general relativity. It is then clear that quantization does not spoil general covariance and
that all quantum effects respect this symmetry. The fluctuation of the metric around the background is the graviton
\begin{equation}
g_{\mu\nu}(x) = \bar{g}_{\mu\nu}(x)+ \kappa h_{\mu\nu}(x)
\end{equation}
and the action can be expanded in powers of $h_{\mu\nu}(x)$ (with corresponding powers of $\kappa$). The Feynman rules
after gauge fixing and the addition of ghosts have been given in several places\cite{Donoghue:1995cz, Donoghue:1994dn, 'tHooft:1974bx}
and need not be repeated here. They are unremarkable aside from the complexity of the tensor indices involved.

Renormalization also proceeds straightforwardly. As advertised, the divergences are local, with the one loop effect being
equivalent to\cite{'tHooft:1974bx}
\begin{equation}
\Delta {\cal L} =\frac{1}{16\pi^2}\frac{2}{4-d} \left[\frac{1}{120}R^2 + \frac{7}{20} R_{\mu\nu}R^{\mu\nu}\right]
\end{equation}
The divergence can then be easily absorbed into renormalized values of the coefficients $c_{1,2}$. The fact
that these occur at the order of four derivatives can be seen by counting powers of $\kappa$ and is completely
equivalent to the energy expansion of the linear sigma model described above.

Pure gravity is one-loop finite. This is because the equations of motion for pure gravity (not including the
cosmological constant) is $R_{\mu\nu}=0$, so that both of the divergent counter terms vanish when treated as a
perturbation. This is an interesting and useful fact, although in the real world it does not imply any special
finiteness to the theory because in the presence of matter the counterterms are physically relevant.

\subsection{Predictions of the effective field theory}

Perhaps the most elementary prediction of quantum general relativity is the scattering of
two gravitons. This was worked out to one-loop by Dunbar and Norridge \cite{Dunbar:1994bn}. The form is
\begin{eqnarray}\label{eq:2}
{\cal A}(++;++) & = &{i\over4}\,{\kappa^2 s^3 \over t u}\left(1 + \frac{\kappa^2~s~t~u }
{4(4\pi)^{2-\epsilon}}\,
 \frac{\Gamma^2(1-\epsilon)\Gamma(1+\epsilon)}
 {\Gamma(1-2\epsilon)}\,  ~\times
 \right.\nonumber \\
&&\left.\hspace{-0em}\times\left[\rule{0pt}{4.5ex}\right.
\frac{2}{\epsilon}\left(
\frac{\ln(-u)}{st}\,+\,\frac{\ln(-t)}{su}\,+\,\frac{\ln(-s)}{tu}
\right)+\,\frac{1}{s^2}\,f\left(\frac{-t}{s},\frac{-u}{s}\right)\right.
\nonumber\\
&&\left.\hspace{1.4em}
+2\,\left(\frac{\ln(-u)\ln(-s)}{su}\,+\,\frac{\ln(-t)\ln(-s)}{tu}\,+\,
\frac{\ln(-t)\ln(-s)}{ts}\right)
\left.\rule{0pt}{4.5ex}\right]\right)\nonumber \\
{\cal A}(++;--) & = & -i\,{\kappa^4 \over 30720
\pi^2}
\left( s^2+t^2 + u^2 \right)   \nonumber\\
{\cal A}(++;+-) & = & -{1 \over 3}
{\cal A}(++;--)
\end{eqnarray}
where
\begin{eqnarray}\label{eq:f}
f\left(\frac{-t}{s},\frac{-u}{s}\right)&=&
\frac{(t+2u)(2t+u)\left(2t^4+2t^3u-t^2u^2+2tu^3+2u^4\right)}
{s^6}
\left(\ln^2\frac{t}{u}+\pi^2\right)\nonumber\\&&
+\frac{(t-u)\left(341t^4+1609t^3u+2566t^2u^2+1609tu^3+
341u^4\right)}
{30s^5}\ln\frac{t}{u}\nonumber\\&&
+\frac{1922t^4+9143t^3u+14622t^2u^2+9143tu^3+1922u^4}
{180s^4},
\end{eqnarray}
and where the $+,~ -$ refer to the graviton helicities. Here ${\cal A}(++;++)$ shows the nature of the energy expansion for general
relativity most clearly. It is
the only amplitude with a tree level matrix element - the others all vanish at tree level. The tree amplitude is corrected by terms
at the next order in the energy expansion, i.e. by factors of order $\kappa^2 ({\rm Energy})^2$ relative to the leading contribution.
Note also the nonanalytic logarithms. Another
interesting feature, despite the presence of $1/\epsilon$ terms in the formulas, is that these results are finite without any
unknown parameters. Because the counterterms vanish in pure gravity, as noted above,
the scattering amplitudes cannot depend on the coefficients of the higher order terms.
The $1/\epsilon$ terms have been shown to be totally infrared in origin, and are canceled
as usual by the inclusion of gravitational bremsstrahlung \cite{Donoghue:1996mt}, as would be
expected from general principles\cite{Weinberg:1965nx}.
This result is a beautiful ``low energy theorem'' of quantum gravity. No matter what
the ultimate ultraviolet completion of the gravitational theory, the scattering process must have this form and
only this form, with no free parameters, as long as the full theory limits to general relativity at low energy.

The leading quantum correction to the gravitational potential is also a low energy theorem independent of
the ultimate high energy theory. The result for the potential of gravitational scattering of two heavy masses is\cite{potential, kk}
\begin{equation}
V(r) = -{GMm \over r} \left[ 1 + 3 {G(M+m)\over rc^2} + \frac{41}{10\pi}
{G\hbar\over r^2 c^3}\right]    \ \ .
\label{potential}
\end{equation}
where the last term is the quantum correction and the term preceding that is the classical
post-Newtonian correction. Let me explain how this is calculated and why it is reliable.
We already know the form of the quantum correction from dimensional analysis, as the unique
dimensionless parameter linear in $\hbar$ and linear in $G$ is $G\hbar/r^2c^3$. The classical post-Newtonian correction
is also a well-known dimensionless combination, without $\hbar$. Fourier transforming tells us that the corresponding
results in momentum space are
\begin{equation}
   \frac{1}{r} \sim \frac{1}{q^2} , ~~~~~~~\frac{1}{r^2} \sim \frac{1}{q^2} \times \sqrt{q^2},~~~~~~~\frac{1}{r^3}
   \sim \frac{1}{q^2}\times q^2\ln q^2, ~~~~~~~  \delta^3(\mathbf{x}) \sim  \frac{1}{q^2}\times q^2
\end{equation}
So here we see the nature of the energy expansion. The leading potential comes from one-graviton exchange with
the $1/q^2$ coming from the massless propagator. The corrections come from loop diagrams - all the one-loop
diagrams that can contribute to the scattering of two masses. The kinematic dependence of the loops then brings
in nonanalytic corrections of the form $Gm\sqrt{q^2}$, $Gq^2 \ln q^2$,  as well as analytic terms $Gq^2$. As shown
above in the chiral scattering result, the analytic terms include the effects of the next order
Lagrangian (the $c_i$ coefficients from Eq. \ref{gravL}) and the divergences.
However, since $\frac{1}{q^2}\times q^2 \sim {\rm constant}$, the Fourier transforms of the
analytic term appears as a delta function in position space, and these terms do not lead to any
long-distance modifications of the potential. The power-law corrections come from the non-analytic
terms, with the logarithm generating the quantum correction. This standard reasoning of effective field
theory allows us to know in advance that the quantum correction will be finite and free from unknown parameters.

The actual calculation is pretty standard, although it is notationally complicated because of all the
tensor indices for graviton couplings. After a series of partial calculations, possible mistakes and
alternative definitions\cite{otherpotentials}, the result appears solid with two groups in
agreement\footnote{I also know of some unpublished confirmations of the same result.}\cite{potential,kk}. However,
the result itself is interesting mainly because of the understanding of why it is calculable. In this calculation
we see the compatibility of general relativity and quantum mechanics at low energy, and have separated off the unknown
high-energy portion of the theory. The magnitude of the correction is far too small to be seen - a correction of $10^{-40}$
at a distance of one fermi. There are no free parameters that we can adjust to change this fact. However, in some ways
this can be pitched as a positive. When we do perturbation theory, the calculations are the most reliable when the
corrections are small. The gravitational quantum correction is the smallest perturbative correction
of all our fundamental theories. So instead of
general relativity being the worst quantum theory as is normally advertised, perhaps it should be considered the best!

It is also worth commenting on the classical correction in Eq. \ref{potential}\footnote{The classical correction was previously known to be
calculable using field theory methods \cite{Iwasaki:1971vb, Gupta:1980zu}.}. This comes from a one-loop
calculation, which surprises some who know the supposed theorem that the loop expansion is an
expansion in $\hbar$. However, that theorem is not really a theorem, and this is one of several
counterexamples\cite{Holstein:2004dn}. The usual proof neglects to note that there also can be
a factor of $\hbar$ within the Lagrangian, as mass terms carry a factor of $m/\hbar$. This is
manifest within the calculation in the square-root nonanalytic term as an extra factor $\sqrt{m^2/k^2\hbar^2}$.

This is perhaps a good place to note that a powerful {\em classical} EFT treatment of gravity has been
developed by Goldberger, Rothstein and their collaborators\cite{Goldberger:2004jt}. This requires a further
development of EFT methods to separate out the relevant components within the graviton field itself. Their
work provides a systematic treatment of the classical bound state and gravitational radiation problems.
It is now a useful component of the gravity wave community, presently ahead of conventional methods in
the prediction of gravity waves of binaries with spin.

Another calculation that shows how EFT methods work is the quantum correction to the Reissner-Nordstom
and Kerr-Newman metrics (for charged objects without and with spin), which takes the form\cite{Donoghue:2001qc} in harmonic gauge
\begin{eqnarray}
   g_{00}&=& 1-{2Gm\over r}+{G\alpha\over r^2}
  - {8G\alpha\hbar \over 3\pi mr^3} +\ldots\nonumber\\
   g_{0i}&=&({2G\over r^3}-{G\alpha\over
   mr^4}+{2G\alpha\hbar\over \pi m^2r^5})(\vec{S}\times\vec{r})_i\nonumber\\
   g_{ij}&=&-\delta_{ij}-\delta_{ij}{2Gm\over r}+G\alpha{r_ir_j\over
   r^2}+ {4G\alpha\hbar\over 3\pi mr^3}\left({r_ir_j\over
   r^2}-\delta_{ij}\right)+\ldots
   \end{eqnarray}
Here the quantum correction comes from photon loops, and gravity is treated classically. But the techniques
are the same. One finds that the classical correction again comes from the square-root non-analytic term in
momentum space. Physically we can identify exactly what causes this correction. It comes from the electric
field which surrounds the charged object. The electric field is non-local, falling with $r^{-2}$ and it carries energy.
Gravity couples to this energy and precisely\cite{Donoghue:2001qc} reproduces the classical correction in the metric.
At tree-level in field theory one sees only the point charged particle, and
the photon loop diagram then is needed to describe the energy in
the electric field surrounding the charged object. The quantum correction again comes from the logarithm in momentum space.

A calculation of Hawking radiation that appears solidly in the spirit of effective field theory is by
Burgess and Hambli\cite{Hambli:1995pp}. They study a scalar propagator in a low curvature region outside
the black hole and regularize with a high energy cutoff. The flux from the black hole can calculated from
the propagator and appears insensitive to the cutoff. However, EFT cannot address questions about the end
state of black hole evaporation - this concerns a situation beyond the region of validity of the effective field theory.

\section{Issues in the gravitational effective field theory}

Effective field theories are expected to have limits. While it is logically conceivable that we could find some
innovative method that allows one to extend general relativity to all energies, this is generally viewed as
unlikely. More typically, effective field theories get modified by new particles and new interactions as one
goes to high energy. The archetypical example is QCD, where the pions of the effective theory get replaced by
the quarks and gluons of QCD at high energy. The most common expectation is that something similar happens for
general relativity. String theory would be a consistent example for general relativity, although there may be
other ways to form an ultraviolet completion of the gravitational interactions. It would be lovely to understand
the nature of this high energy modification. However, since physics is an experimental science and the Planck scale
appears to be frustratingly out of reach of experiment, it is not clear that we will have reliable evidence for the
nature of the ultimate theory in the foreseeable future. The competition between divergent proposals for the ultimate
theory, which in principle would be a scientific discussion if experiment could keep up, may remain unresolved in our
lifetimes without the input of experiment.

However, it does remain interesting to explore the limits of the effective theory. Potentially these
limits can help us understand when new physics
must enter. One obvious case is at the Planck energy. Scattering amplitudes are all proportional to powers of $GE^2$, with
\begin{equation}
{\cal M}= {\cal M}_0 [1 + a G E^2 +b G E^2 \ln E^2+.... ]
\end{equation}
This clearly leads to problems at the Planck scale, where the energy expansion breaks down.

However, there may also be other limits to the validity of the effective theory. I have argued
elsewhere\cite{Donoghue:1994dn, Donoghue:2009mn} that the extreme infrared of general relativity
may pose problems that other effective theories do not face. This is because gravitational
effects may build up - the integrated curvature may be large even in the local curvature is not.
Such effects are manifest as horizons and singularities. Horizons by themselves are not expected to
be problematic - in a local neighborhood nothing special need be evident. But the long distance
relations of horizons to spatial infinity clearly brings in problematic features for black holes.

Moreover, singularities pose problems for the effective theory in the long distance limit. The
singularity itself is not the problem - we know that the effective theory breaks down when the
curvature is large. But even for the long distance theory, propagating past a singularity poses
some difficulties. We do not know the fate of the modes that flow into the singularity. Perhaps
this can be solved by treating the location of the singularity as an external source. In chiral
perturbation theory, this is done for baryons, which arguably appear as solitons in the effective
theory which are found only beyond the limits of the effective theory, yet also serve as a source
for the Goldstone bosons. Perhaps singular regions can be treated as localized sources of gravitons.

However, even if singularities can be isolated and tamed, there are issues of the relation of
localized regions of small curvature, where the effective theory is demonstrably valid, to far
distant regions where the integrated curvature is large. This conflict is at the heart of black
hole paradoxes, where the horizon is well behave locally but is problematic when defined by its relation to spatial infinity.

\section{Gravitational corrections and running coupling constants}

Let me also address what effective field theory has to say about the gravitational corrections
to the running of coupling constants, including that of gravity itself. This subject has had
a confusing recent history, and EFT is useful in sorting out the issues\footnote{The comments of
this section continue to assume that the cosmological constant is small enough to be neglected.
In the presence of a cosmological constant the story is different and there can be genuine contributions to running couplings\cite{Toms:2009vd}}.

Despite having a pre-history\cite{Kiritsis:1994ta},
recent activity stems from the work of Robinson and Wilczek\cite{Robinson:2005fj}, who suggested that
the beta function of a gauge theory could have the form
\begin{equation}
\beta (e, E) = \frac{b_0}{(4\pi)^2}e^3 + a_0 e\kappa^2E^2
\end{equation}
and calculated $a_0$ to be negative. While this correction is tiny for most energies,
the negative sign suggests that all couplings could be asymptotically free if naively
extrapolated past the Planck scale. Subsequent work by several authors, all using
dimensional regularization, found that the gravitational correction to the running coupling
vanishes\cite{dimreg}. Further work, including some of the same authors, using variations of
cutoff ($\Lambda$) regularization then found that it does run\cite{cutoff}, where the cutoff
plays the role of the energy, i.e. $\beta \sim G\Lambda^2$. Papers trying to
clarify this muddle include\cite{ad1, ad2, Toms:2011zza, Ellis:2010rw}. My treatment
here most naturally follows the ones of my collaborators and myself\cite{ad1,ad2}

In renormalizable field theories, the use of a running coupling is both useful and universal.
It is useful because it sums up a set of quantum corrections which potentially could have large
logarithmic factors. It is universal because the logarithmic factors are tied to the renormalization
of the charge, and hence enter the same way in all processes. Specifically, when one employs
dimensional regularization the charge renormalization and the $\ln \mu^2$ factors always enter the same way, because of the expansion
of $(\mu^2)^\epsilon/\epsilon$. For example, in perturbation theory photon exchange at high energy involves
\begin{equation}
\frac{e^2}{q^2} =\frac{e_0^2}{q^2}\left[1+ \frac{\alpha}{3\pi^2}\left(\frac{2}{4-d} - \ln\frac{-q^2}{\mu^2}+...\right)\right]
\end{equation}
The $\ln \mu^2$ dependence follows the renormalization of the charge, and dimensionally the $\ln q^2$ dependence has to accompany
the
$\ln \mu^2 $ dependence. When this is turned into a running coupling, the energy dependence always appears in a universal
fashion because the charge is renormalized the same way in all processes.
In addition, with logarithmic running the running coupling constant is crossing symmetric
since $\ln q^2$ has the same value, up to an imaginary part due to on-shell intermediate states,
for $q^2$ either positive or negative. This lets the running coupling constant be applicable in
both the direct channel and in the crossed channel for a given type of reaction.

For gravitational corrections in the perturbative regime, effective field theory explains why each
of the features (useful, universal, crossing symmetric) no longer holds. Let me discuss them individually.

We have seen that gravitational corrections are expansions in $Gq^2$ with potentially extra
logarithms. Because in different reactions $q^2$ can take on either sign, the gravitational
correction will not have the same sign under crossing. For example if we calculated the
gravitational correction to a process such as $f+ \bar{f} \to f'+\bar{f}'$, for two different
flavors $f,~f'$, within QED, one would find a correction of the form
\begin{equation}
{\cal M} \sim \frac{e^2}{s}[1 + bGs]
\end{equation}
with $s=(p_1+p_2)^2>0$ and $b$ is a constant to be calculated. However, if one studies the
related crossed process $f +f' \to f+f'$ one would have
\begin{equation}
{\cal M} \sim \frac{e^2}{t}[1 + bGt]
\end{equation}
with $t=(p_1-p_3)^2<0$ and $b$ is the same constant. If one tried to absorb the quantum correction
into a running coupling, it would be an increasing function of energy in the one process and a
decreasing function of the energy in the other process. Moreover, other processes such as the single
flavor case $f+ \bar{f} \to f+\bar{f}$, would involve both $s$ and $t$ variables within the same reaction.
A universal energy dependent running coupling with quadratic four-momentum dependence cannot account for the features of crossing.

Effective field theory also explains why gravitational corrections are not a universal factor correcting
the coupling. We have seen in the direct calculations above and in the discussion of the energy expansion
that gravitational corrections do not renormalize the original operator, but generate divergences that are
two factors of the energy (derivatives) higher. This of course is due to the dimensional coupling constant.
For gravity itself, after starting with the Einstein action $R$ we found that divergences were proportional
to $R^2$ and $R_{\mu\nu}R^{\mu\nu}$. For the QED case, the higher order operators could be
\begin{equation}
\bar{\psi}\gamma_\mu\partial_\nu \psi \partial^\mu A^\nu,~~~~~~~~~~~~\bar{\psi}\gamma_\mu\psi \partial^2 A^\mu, ~~~~~~~~~~~~\partial^\mu F^{\lambda\nu}\partial_\lambda F_{\mu\nu}, ~~~~
~~~~~~~~\bar{\psi}\gamma_\mu\psi \bar{\psi}\gamma^\mu\psi ~~.
\end{equation}
These operators are related to each other by the equations of motion, but there are two points associated
with this observation. One is that, in contrast to renormalizable field theories, we are
not renormalizing the original coupling so that we do not have any expectation of universality. But in addition,
different processes involve different combinations of the higher order operators, so that the renormalization
that takes place is intrinsically different for different processes. Different reactions will involve
different factors. It is for this reason that use of the renormalization group in effective field theory,
which is in fact a well studied subject\cite{Weinberg:1978kz, rg}, does not involve the running of the basic coupling, but instead
is limited to predicting logarithmic factors associated with the higher order operators.

So we have seen that effective field theory, in the region where we have control over the calculation,
does not lead to gravitational corrections to the original running couplings. We might still ask if there might
nevertheless be some useful definition that plays such a role. This logically could occur if one
repackaged some of the effects of the higher order operators {\em as if} they were the original
operator - i.e. some sort of truncation of the operator basis. However, here one can simply calculate
gravitational corrections to various processes to see if some useful definition emerges. The answer is
decidedly negative\cite{ad2}. The numerical factors involved in different reactions vary by large factors and come with both signs.

In the perturbative regime there is no useful and universal definition of running $G(q^2)$ nor gravitational
corrections to running $e(q^2)$, and there is a kinematic crossing obstacle to any conceivable proposal.

Far too often in the literature, a dimensional cutoff ($\Lambda$) is employed to incorrectly conclude that
it contributes to a running coupling. For example, one can use cutoff and find that the original coupling
{\em is} renormalized. For example for gravity or QED
\begin{equation}
G = G_0[1+cG_0\Lambda^2], ~~~~~~~e^2= e_0^2[1+dG\Lambda^2]
\end{equation}
for some constants $c, ~d$. If one then says the magic phrase ``Wilsonian'', one might think that
this then defines a running coupling $G(\Lambda)$ or $e(\Lambda)$. However, this is not a running coupling -
it is just renormalization. The quadratic $\Lambda$ dependence disappears into the renormalized value
of $G$ or $e$ at low energy\cite{ad2, Toms:2011zza}. Once you measure this value there is no remaining
quadratic $\Lambda$ dependence, and there is no energy dependence that tracks the $G\Lambda^2$ behavior.

These observations pose questions for the sub-field of Asymptotic Safety\cite{Weinberg:1980gg, asymsafety}.
In this area, on considers a Euclidean coupling $g=Gk_E^2$, where $k_E$ is a Euclidean momenta, and attempts
to find a ultraviolet fixed point for $g$. Since the operator basis expands to operators of increasingly high
number of derivatives, this involves a truncation of the operator basis. Given such a Euclidean truncation,
one can by construction define a running coupling, and one does find a UV fixed point well beyond the Planck
scale. However, the question remains whether this implies anything useful for real processes in Lorentzian gravity.
One would think that one should find evidence of such a running coupling in the perturbative region where we have
control over the calculation. However, we have seen the reasons why this does not occur. The conflict certainly deserves further study.

\section{Summary}

Effective field theory is a well-developed framework for isolating the quantum effects of low energy particles and
interactions, even if these interactions by themselves fall outside of a complete renormalizable field theory. It
works well with general relativity applied over ordinary curvatures and energy scales. The effective field theory
has limits to its validity, most notably it is limited to scales below the Planck energy, and does not resolve all
of the issues of quantum gravity. However, effective field theory has shown that general relativity and quantum
mechanics do in fact go together fine at ordinary scales where both are valid. GR behaves like an ordinary field
theory over those scales. This is important progress. We still have work to do in order to understand
gravity and the other interactions at extreme scales.


\begin{theacknowledgments}
This material was presented at the Sixth International School on Field Theory and Gravitation, Petropolis Brazil, April 2012.
I thank the organizers and my fellow participants for an interesting and informative school. This work is supported in
part by the U.S. National Science Foundation grant PHY-0855119
\end{theacknowledgments}



\bibliographystyle{aipproc}   




\end{document}